\documentclass[preprint,showpacs,preprintnumbers,amsmath,amssymb]{revtex4}

\usepackage{graphicx}% Include figure files
\usepackage{dcolumn}% Align table columns on decimal point
\usepackage{bm}% bold math

\begin{document}

\title{Imperfections in Focal Conic Domains: the Role of Dislocations}
\author{M. Kleman$^{1}$}\email{maurice.kleman@mines.org}
\author {C. Meyer$^{2}$}
\author {Yu. A. Nastishin$^{1,3}$}
\affiliation{$^{1}$Institut de Min\'{e}ralogie et de Physique des
Milieux Condens\'{e}s (UMR CNRS 7590), Universit\'{e}
Pierre-et-Marie-Curie, Campus Boucicaut, 140 rue de Lourmel, 75015
Paris, France\\
$^{2}$Laboratoire de Physique de la Mati\`{e}re Condens\'{e}e,
Universit\'{e}
\\
de Picardie, 33 rue Saint-Leu, 80039 Amiens, France\\
$^{3}$Institute for Physical Optics, 23 Dragomanov str. Lviv 79005 Ukraine}

\date{\today}

\begin{abstract}
It is usual to think of Focal Conic Domains (FCD) as perfect
geometric constructions in which the layers are folded into Dupin
cyclides, about an ellipse and a hyperbola that are conjugate.
This ideal picture is often far from reality. We have investigated
in detail  the FCDs in several materials that have a transition
from a smectic A (SmA) to a nematic phase (N). The ellipse and the
hyperbola are seldom perfect, and the FCD textures also suffer
large transformations (in shape or/and in nature) when approaching
the transition to the nematic phase, or appear imperfect on
cooling from the nematic phase.  We interpret these imperfections
as due to the interaction of FCDs with dislocations.  We analyze
theoretically the general principles subtending the interaction
mechanisms between FCDs and finite Burgers vector dislocations,
namely the formation of \textit{kinks} on disclinations, to which
dislocations are attached, and we present models relating to some
experimental results. Whereas the principles of the interactions
are very general, their realizations can differ widely in function
of the boundary conditions.
\end{abstract}

\maketitle

\section{Introduction}
\label{sec:1} The discussion that follows, about the behavior of
defects in the SmA (smectic A) phase is inspired by a few
experimental polarized light microscopy observations reported in
\cite{meyer1} and summarized below.  These observations have since
been developed \cite{yuriy}.  They relate to a domain of
temperature that extends approximately $1^{\circ}C$  below the
SmA$\longrightarrow$N phase transition, some of the most relevant
experiments having been done with an accuracy of $\pm 1mK$. The
very near vicinity of the transition, where phenomena usually
qualified of transitional do happen, could not be investigated,
and then has not been. It appears in the domain we have searched,
the focal conic domains suffer considerable visible modifications,
with remarkable imperfections in shape. It is the nature of these
interactions that we wish to describe in the present article.

The defects and textures of the SmA and N phases are reasonably
well understood at mesoscopic and macroscopic scales, at least for
their static physical and topological properties. Contrariwise,
the role played by the smectic defects at the phase transition has
been little investigated.  It is precisely in this region that the
FCDs (focal conic domains), the only defects that are fully
observable in light microscopy, show these large modifications
that, we shall argue, are essentially due to their interactions
with dislocations.  The SmA$\longrightarrow$N phase transition has
been the object of many investigations (for a review, see
\cite{degennes}). The compression modulus $B$ tends towards a
value (equal to or slightly different from zero, according to the
author, see \textit{e.g.},
\cite{benzekri,rogez,beaubois,yethiraj}); its variation is
noticeable in a large temperature range (more than half a degree
in the compounds that we have investigated). Notice that in this
range, $K_1$ (the splay modulus) stays practically constant. The
question of $\overline{K}$ (the saddle-splay modulus) has been
little investigated yet, either theoretically or experimentally
(see \cite{barbero} for the nematic phase); the results that
follow have been interpreted by assuming that $\overline{K}$ too
stays practically constant. $K_2$ (the twist modulus) is infinite
in layered media as long as the director remains along the layer
normal (which might be infirmed very close to the nematic
transition).\\
Let us now recall some defect features of the SmA phase.  These
defects are of two types, focal conic domains (which are especial
types of disclinations) and
dislocations:\\
\noindent - \textit{focal conic domains (FCDs)}: the layers are
parallel, so that there is no strain energy but only curvature
energy.  The normals to the layers envelop two focal surfaces on
which the curvature is infinite (the energy diverges). The focal
surfaces are degenerate into lines in order to minimize this
large curvature energy.  These lines are necessarily two
confocal conics, an ellipse and a hyperbola, observable by
optical microscopy
\cite{friedela,friedelb,klemanbook,williams0,fournier}. The
layers are folded along \textit{Dupin cyclides}, surfaces that
have the topology of tori. And indeed the simplest geometric
case is when the ellipse E is degenerate into a circle - the
confocal hyperbola H being degenerate into a straight line
perpendicular to the plane of the circle and going through its
centre. In this case the layers are nested tori, restricted in
fact to those parts of the tori that have negative Gaussian
curvature $G=\sigma_1 \sigma_2$. The $G<0$ case is indeed the
most usual case met experimentally in generic Dupin cyclides,
see \cite{maurice,boltenhagen1}. We shall not consider in the
sequel the situations where the layers are restricted to those
parts that have positive Gaussian curvature; and as a matter of
fact the mixed case is not observed. In the toric case just
alluded, the focal conic \textit{domain} is the region of space
occupied by those nested layers restricted to their $G<0$ parts;
it is bound by a cylinder parallel to H and whose cross section
is E. In the generic case, the region of space where the layers
have $G<0$ is bound by two half-cylinders of revolution, that
meet on the ellipse, and whose generatrices are parallel to the
hyperbola asymptotes Fig.1a. This is the picture of an ideal,
\textit{complete}, FCD. Fig. 1b illustrates a case where $G<0$
and $G>0$ regions are visible in the same FCD; it does not
correspond to any situation met in practice. Models for
\textit{incomplete} FCDs are shown farther ahead (Fig.9a and
9b). The important question how FCDs are packed in space
\cite{friedelb,klemanbook} will be approached, but only
incidentally.

The curvature energy $f_{FCD}$ of a complete, ideal, focal conic
domain depends on $K_1$ and $\bar{K}$:

\begin{equation}
f_{FCD}=f_{bulk}+f_{core}=4\pi a (1-e^2)K(e^2)[K_1 \ln
\displaystyle \frac{2b}{\xi}-2K_1-\bar{K}]\\
 +f_{core}\\
\end{equation}
%eq.1
where $a$ is the semi-major axis of the ellipse, $b$ the
semi-minor axis, $e$ the eccentricity and $K(e^2)$ the complete
elliptic integral of the first species \cite{maurice}. It is
believed that the energy $f_{strain}$ attached to the thickness
variation of the layers is negligible compared to $f_{FCD}$.
Very little is known about the core contribution $f_{core}$, but
it is usually assumed that it scales as $a K_1$. Thus, at $a$
and $e$ constant, the FCD total energy does not vary
significantly in the domain of temperature under investigation,
if our assumptions about the temperature variation of $K_1$ and
$\bar{K}$ turn to be
true.\\
\noindent - \textit{screw dislocation lines and edge dislocation
lines:} : their unit length line energies can be written:\\
\begin{equation}
\label{dislenergy}
 f_s=\displaystyle \frac{1}{128} \frac{B
b_{disl}^4}{r_{c, screw}^2}+f_{core},  f_{e}=\displaystyle
\frac{1}{2} \sqrt{K_1 B} \displaystyle
\frac{b_{disl}^2}{r_{c,edge}}+f_{core},
\end{equation}
%eq.2
where $b_{disl}=n d_0$ is the dislocation Burgers vector ($d_0$
is the layer thickness) and $r_c$ is the core radius. It is
visible that the elastic contributions (the off-core terms in
Eq. \ref{dislenergy}) decrease when $T$ gets closer to $T_{AN}$,
because $B$ decreases whereas the temperature changes of the
core energies can be neglected or even decrease - indeed, in a
na\"{\i}ve model inspired by the solid-liquid transition, these
energies are of order $k_B(T_{AN}-T) \displaystyle \frac{\pi
r_c^2}{\delta^2 d_0}$ per unit length of dislocation line,
\textit{i.e.} small; $\delta^2 d_0$ being the volume occupied by
a molecule, and $r_c$ is microscopic, practically constant (of
order $\delta$ for a screw dislocation, $d_0$ for an edge
dislocation). The decrease of $B$, as already stated, is
effective on a large temperature range before the transition,
probably larger than $1^{\circ}C$, see
\cite{benzekri,rogez,beaubois}. The core radii scale as the
correlation lengths very close to
the transition, but this region is of no interest to us.\\

\noindent \textit{Comments on the experimental conditions}\\
 The FCDs are stable and immobile  in the lower part
of the temperature range we have investigated, they however quite
often display variations to their ideal shape. The transformations
of the FCDs, when approaching the transition, are visible with a
simple optical microscopy set up. They appear as rather sudden
phenomena, about one degree before reaching $T_{AN}$ from below,
at a temperature $T^*$ that depends slightly on the boundary
conditions. Our observations relate to \textit{thermotropic}
compounds, two belonging to the cyanobiphenyl series, 8CB and 9CB.
We also noticed that the FCD texture of  the 10CB compound does
not display any conspicuous modifications when $T$ increases, but
this chemical does not have a nematic phase; the transition is
direct to the isotropic phase. This is in contrast with the other
compounds, that have a SmA$\rightarrow$N transition. In these
latter cases, either the FCDs disappear by shrinking before the
phase transition, or the ellipse and the hyperbola transform into
disclinations in the nematic phase; the first situation occurs
usually for small and medium size (tens of microns in diameter)
slowly FCDs heated (heating rate lower than $0.1^{\circ}C/min$),
the second one occurs for large (hundred of microns and more in
diameter) FCDs, when they are brought to the transition under
faster heating (this linear scaling is conditional, depending on
the heating rate: at a high heating rate of several degrees per
minute, even small FCDs have no time to shrink). When cooling down
from the nematic phase, the FCD textures in 8CB, 8OCB and 9CB
usually do not display ideal FCDs. Instead FCD fragments grow,
join and form focal domains, which in many cases are not ideal.
The double helical objects described in \cite{williams1}, which
are splitting modes of giant screw dislocations, are obtained this
way. These \textit{imperfect} FCDs (iFCD) can be quenched to lower
temperatures where they stabilize because of a much reduced
mobility (high viscosity). The boundary conditions play an
important role in the definition of the final texture.\\
We believe that the transformations of the FCD texture in 8CB,
9CB, when approaching the nematic phase, as well as the formation
of iFCDs when coming from above, are due to the interaction of the
FCDs with dislocations. Dislocations are generally not visible by
optical microscopy, except when their Burgers vector is large
(micron size), which situation occurs for edge dislocations,
clustering into oily streaks \cite{friedela,friedelb,klemanbook}
or screw dislocations split into two $k=\frac{1}{2}$ disclinations
\cite{williams1,kln}. We argue here that the presence of numerous
dislocations can be revealed via their distorting action on the
FCDs, which are visible.

\section{GEOMETRIC RULES FOR IDEAL FOCAL CONIC DOMAINS}
Essential for a better understanding of the modifications suffered
by FCDs when interacting with dislocations are the following
properties, that characterize them when they are in an
\textit{ideal} state.

(a)- Projected orthogonally upon a plane, along any direction, the
ellipse E and the hyperbola H cross at right angles, Fig.2a. This
is a particular case of Darboux's theorem \cite{darboux}, which
states that if a congruence of straight lines is orthogonal to a
set of parallel surfaces, the two focal surfaces $\Sigma_1$ and
$\Sigma_2$ (that this congruence generically envelops) are such
that the planes tangent to $\Sigma_1$ and $\Sigma_2$ at the
contact points $M_1$ and $M_2$ of any line $\Delta$ of the
congruence are orthogonal. This is the reason why the projections
of the ellipse E and the hyperbola H belonging to the same FCD
look orthogonal. Here the straight line $\Delta$ is a normal to
the Dupin cyclides, and indicates the average direction of the
molecules. Darboux's theorem is empirically satisfied by a number
of FCDs, which in that sense are ideal FCDs; when it is not, (see
Fig.2b) it implies that the FCD in question is geometrically
interacting with other defects, as we shall discuss in the sequel.

(b)- The inner layers of an \textit{isolated complete} FCD can be
continued outside the FCD by planar layers perpendicular to the
asymptotes (this is obvious from Fig. 1a), which can form two sets
of parallel planes meeting on the plane of the ellipse, along a
direction parallel to the minor axis of the ellipse, at an angle
$\omega$. The plane of the ellipse is therefore a tilt grain
boundary. In a solid crystal, a tilt grain boundary is usually
split into edge dislocations whose Burgers vectors are
perpendicular to its plane. The same is of course possible for a
tilt grain boundary in a layered medium. One expects that some of
those dislocations meet the ellipse. As a matter of fact, the
ellipse of an isolated FCD is the termination of a set of
dislocations whose total Burgers vector $b_{disl}=4ae=4c$, as
explained below.

(c)- Two neighboring ideal FCDs whose ellipses are in the same
plane and tangent at some point M are in contact along at least
one line segment joining M to a point P at which the two
hyperbolae intersect. This geometry, frequently observed, is a
particular realization of the law of corresponding cones
\cite{friedela,friedelb,klemanbook}, a geometrical property that
rules the way FCDs pack in space. A tilt grain boundary whose
angle of misorientation $\omega$ is neither too small nor too
large is usually made of a FCD packing such that the ellipses
belong to the grain boundary, have a constant eccentricity $e =
\sin\frac{\omega}{2}$, the asymptotes of the hyperbolae being
parallel \cite{kleman1}, see Fig.3. And indeed the free
interstices of the packing of ellipses in the plane of the grain
boundary are filled with dislocations \cite{kleman1}. There is
therefore a relation of equivalence between dislocations and focal
conic domains \cite{bourdon, boltenhagen2}.

\section{KINKS ON DISCLINATIONS}
\subsection{Wedge and twist disclinations. FCD confocal conics are disclinations.}

Disclinations are typical line defects in a medium endowed with
a director order parameter \cite{frank}. One distinguishes
\textit{wedge} disclinations, whose rotation vector
$\overrightarrow{\Omega}$ is along the disclination line, and
\textit{twist} disclinations, whose rotation vector
$\overrightarrow{\Omega}$ is orthogonal to the disclination
line. As shown in \cite{friedela,friedelb}, there are
necessarily dislocations attached to a line segment of twist
character.  Let us remind that the focal lines of a FCD, are, by
nature, disclinations.

(a)- The hyperbola is a disclination of strength $k=1$, whose
rotation vector ($\overrightarrow{\Omega}=2\pi \overrightarrow{t}
$) varies in direction (not in length) all along the hyperbola: at
each point of the hyperbola it is parallel to the tangent
$\overrightarrow{t}$ at this point. The layer geometry is
axial-symmetric in the vicinity of the hyperbola. Insofar as it is
a disclination, the hyperbola is of wedge character; there are no
attached dislocations.

(b)- If the full cyclides are considered, as in Fig. 1b, the
layers surround the ellipse are axial-symmetric about the tangent
to the ellipse; thus the ellipse appears also as a $k=1$ wedge
line. But, if one restricts to the inner $G<0$ layers of the
complete FCD, supplemented by the outer planar layers, the ellipse
appears as a disclination of strength $k=\displaystyle
\frac{1}{2}$, whose rotation vector ($\overrightarrow{\Omega}=\pi
\overrightarrow{t}$) varies (in direction but not in length) all
along the ellipse; $\overrightarrow{t}$ is in the plane of the
ellipse and tangent to the layer inside the ellipse, Fig.4. This
disclination is of mixed character, the attached dislocations are
precisely those that form the tilt boundary
\cite{kleman1,boltenhagen1,friedel1} whose existence has been
established above; see below for details.

\subsection{Kinks, generic properties.}

Modifications to the twist/wedge character of a disclination can
be achieved in the generic case by attaching/detaching new
dislocations to the line. Such operations modify the shape of
the line, by the introduction of \textit{kinks}, Fig.5. For
instance, in order to attach at some point A on a wedge line
$\cal L$ a set of dislocations of total Burgers vector
$\overrightarrow{b}_{disl}$, one has to introduce a kink
$\overrightarrow{AB}$, with a component perpendicular to $\cal
L$, (\textit{i.e.} a segment $\overrightarrow{AB}$ having a
twist component), such that

\begin{equation}
\label{bdislAB}
\overrightarrow{b}_{disl}=2 \sin \displaystyle
\frac {\Omega}{2} \overrightarrow{t} \times \overrightarrow{AB},
\end{equation}
%eq.3

where $\overrightarrow{t}$ is an unit vector tangent to the line
and $\overrightarrow{\Omega}$ ($\overrightarrow{\Omega}=\Omega
\overrightarrow{t}$) is the rotation invariant carried by the
disclination; see \cite{klemanbook,friedel1} and the Appendix for
a demonstration of Eq. (\ref{bdislAB}). In practice lines of
interest are of strength $|k|=\displaystyle \frac{1}{2}$,
$|\overrightarrow{\Omega}|=\pi$. Reciprocally, the presence of a
kink reveals the presence of dislocations attached to the line.
The above picture of a kink says nothing about the nature (edge or
screw) of the attached dislocations, and the way they relax and
disperse through space about the disclination line. The line
flexibility, \textit{i.e.} the main property at work when the
medium is deformed, elastically or by flow, takes its origin here,
in this interplay of the disclination line with dislocations.

A kink can be infinitesimally small;

\begin{equation}
\label{dbdislds}
\overrightarrow{db}_{disl}=2 \sin \displaystyle
\frac {\Omega}{2} \overrightarrow{t} \times \overrightarrow{ds},
\end{equation}
%eq.4

\noindent where $\overrightarrow{ds}$ is an infinitesimal element
along the line \cite{friedel1}. A density of infinitesimally small
kinks modifies the curvature of the line.  A dislocation attached
to an infinitesimally small kink has an infinitesimally small
Burgers vector; a dislocation attached to a finite kink may have a
finite Burgers vector, as we see now.

\subsection{Kinks in a SmA}

Let us now consider in more detail the geometry of the attachment of
dislocations to a focal line in a FCD. We first state some general properties,
and then consider separately the case of the ellipse and the case of the
hyperbola.

Again, the dislocations emanating from the kink have to belong to
one of the two following categories: they are either dislocations
with infinitesimal Burgers vectors whose directions are parallel
to the layer dislocations of the layer stacking, or with Burgers
vectors $|\overrightarrow{b}_{disl}|=n d_0$ perpendicular to the
layer (these are the usual SmA quantified dislocations).  Note
that in both cases the Burgers vectors are translation symmetry
vectors; they are \textit{perfect} Burgers vectors in the sense of
the Volterra process. We consider them successively.

\textit{Infinitesimal Burgers vectors} relate to dislocation
densities that relax by the effect of viscosity; they affect the
curvature of the layers and consequently, as alluded just above,
they also affect their thickness, since the layers have to keep
in contact.  We shall not expatiate on such defects, which are
not relevant to our subject. Just notice that the theory has
been developed for solids since long (see \cite{kroener} for a
general review) and is related to the concept of densities of
infinitesimal dislocations in nematics and cholesterics
introduced first in \cite{friedel1}; see also
\cite{klemanbook,kln}. An essential point worth emphasizing is
that a continuous density of infinitesimal dislocations can be
attended by a strainless, elastically relaxed, state. In our
case, this would correspond to a state where the layers keep
parallel. Continuous dislocation with Burgers vectors parallel
to the layers do not introduce any kind of singularity of the
SmA order parameter. Eq. (\ref{dbdislds}) indicates that the
related kink $\overrightarrow{ds}$ and that $\overrightarrow{t}$
are both perpendicular to $\overrightarrow{db}_{disl}$, which
condition does not specify any special direction for
$d\overrightarrow{s}$.

\textit{Finite Burgers vectors}: this case is better represented
by Eq. (\ref{bdislAB}), because the Burgers vector and the kink
$\overrightarrow{AB}$ are both finite. $\overrightarrow{AB}$ and
$\overrightarrow{t}$ have both to be in a plane tangent to the
local layer. To an elementary dislocation
$|\overrightarrow{b}_{disl}|=d_0$ corresponds an elementary kink.
An elementary kink is microscopic (with $|k|=\displaystyle
\frac{1}{2}$, $|\overrightarrow{\Omega}|=\pi$, one has $AB=2d_0$);
one can thus possibly have a density of such elementary kinks,
rendering the line curved when observed at a mesoscopic scale.
This does not exclude the possibility that infinitesimally small
dislocations are attached to finite kinks.

Simple as they look, the application of these criteria requires
however some care.

\subsection{Quantified Burgers vectors attached to an ellipse.}

Fig.4 is a schematic view of the properties of an ellipse,
belonging to an ideal FCD, which are in relation to its
$k=\displaystyle \frac{1}{2}$ disclination character.  The layer
geometry is different inside and outside the ellipse.
\underline{\textit{Inside}}, the Dupin cyclide layers intersect
the plane of the ellipse \textit{perpendicularly}.
\underline{\textit{Outside}}, the layers are planar and
\textit{perpendicular} to the asymptotic directions. The change of
geometry between the inside and the outside is achieved by a
rotation of the layers about the local rotation vector
$\overrightarrow{\Omega}=\Omega \overrightarrow{t}$;
$\overrightarrow{\Omega}$ is parallel to the layers (inside and
outside) and is along the intersection of the layers with the
plane of the ellipse, inside.

The layer at M (M being a running point on the ellipse) is indeed
folded inside about the local $\overrightarrow{t}$ direction, is
singular at M (it is a conical point), and extends outside along a
fold made of two half planes symmetrical with respect the ellipse
plane, each perpendicular to one or the other of the two
asymptotic directions of the confocal hyperbola, and thereby
making an angle $\omega=2\sin^{-1}e$ about a direction parallel to
the minor axis of the ellipse (see \cite{klemanbook}, chapter 10).
The ellipse plane outside the ellipse is therefore a tilt boundary
of misorientation angle $\omega$, which can be accommodated by
edge dislocations of Burgers vectors multiple of $d_0$,
perpendicular to the plane of the tilt boundary, \textit{i.e.} the
plane of the ellipse outside. There is one such dislocation
$|\overrightarrow{b}_{disl}|=2d_{0}$ per layer counted inside the
ellipse. These results are similar to those obtained in section
II; they also justify the choice of $\overrightarrow{\Omega}$ we
have done for the disclination rotation along the ellipse.

The same result can be obtained by using Eq. (\ref{dbdislds}). Let
us parameterize the ellipse in polar coordinates with the origin
at the physical focus, Fig.6.

\begin{equation}
\label{ellipse}
r=\displaystyle \frac{p}{1+e \cos \phi},
\end{equation}
%eq. 5

\noindent where $p=\displaystyle \frac{b^2}{a}$ and $\phi$ is the
polar angle. Applying Eq. (\ref{dbdislds}), one then finds that
the $k=\displaystyle \frac{1}{2}$ ellipse disclination has an
attached Burgers vector density

\begin{equation}
db_{disl} = 2 dr,
\end{equation}
%eq. 6

The total Burgers vector attached to the ellipse is $\displaystyle
\int_{\phi=0}^{\phi=\pi} db_{disl}=4c$, as indicated above. If one
takes $dr=d_0$, - an approximation which makes sense (up to second
order), since $d_0$ is so small compared to the size $a$ of the
ellipse - it is visible that the points $M\{{r, \phi}\}$ and
$N\{{r + dr, \phi+d\phi}\}$ are on two parallel smectic layers at
a distance $d_0$. Notice that the density of dislocations is
constant if measured along the major axis: $\displaystyle
\frac{db_{disl}}{dx} =-2e$. There are no dislocations attached to
the singular circle of a toric FCD, as the eccentricity e
vanishes. An ellipse can be thought of as a circle kinked at the
layer scale.

\section{KINKED FOCAL CONIC DOMAINS}

\subsection{Frequent geometries for a kinked ellipse.}

The kinking of the ellipse takes different geometries, whether the
dislocations at stake are located inside the ellipse (where
$\overrightarrow{b}_{disl}=nd_{0}$ is in the plane of the ellipse,
the layers being perpendicular to this plane) or outside (where
$\overrightarrow{b}_{disl}=nd_{0}$ is perpendicular to the plane
of the ellipse).\\
\textit{Outside the FCD;} in that case the kinking of the ellipse
is in its plane. This in-plane kinked ellipse, we call it a
Mouse.(Fig.7). If the dislocation lines attached to the ellipse
disperse away \textit{outside} the focal conic domain,
\textit{i.e.} in a region of space where the layers are in the
plane of the ellipse; $\overrightarrow{t}$, which varies in
direction all along the ellipse, is in this plane. Applying Eq.
(\ref{bdislAB}), it appears that the kinks have to be in the plane
of the ellipse. This configuration has been observed, in a
situation where the kinks are so small and have such a high
density that the kinked ellipse appears to be continuous, but its
shape departs considerably from a 'perfect' ellipse; it is
smoothly distorted by the in-plane kinks: that is the reason why
we use the term of Mouse (Fig.7a). Fig.7b provides a model for
such kinks, (which always go by pairs), drawn here at a scale
which has no relation with the real scale. The photograph of
Fig.7a is taken from the rim of a free standing film, in a region
where the thickness of the film is quickly changing, and the wedge
angle $\omega(r)$ between the opposite free boundaries varies
monotonically. The anchoring conditions are homeotropic; there is
therefore a tilt boundary in the mid-plane of the film, but with a
variable misorientation angle. The Mouse is in this mid-plane; the
extra dislocations attached to the kinks (edge dislocations in the
mid-plane) relax the variation of $\omega$ by contributing to the
modification of the density $\displaystyle \frac{db_{disl}}{ds}$
of dislocations in this plane; see \cite{yuriy} for a more
detailed account.

\textit{Inside the FCD}; in that case the kinking of the ellipse
brings a part of it out its plane. This off-plane kinked
ellipse, we call it a Turtle (Fig.8). The layers rotate about
$\overrightarrow{t}$ by an angle of $\pi$; hence they become
perpendicular to the plane of the ellipse, \textit{inside} the
FCD. Therefore the dislocations that disperse away inside are
attached to kinks that are perpendicular to the plane of the
ellipse, on average.  A pair of elementary kinks (not at scale
at all in the figure), symmetric with respect to the major axis,
can be linked by a unique dislocation (Fig.8b). Our observations
(Fig.8a) indicate the existence of another mode of kinking, with
screw dislocations joining the kink (of macroscopic size) to the
hyperbola. There is no kink on the hyperbola, because the two
screw segments are of opposite signs, if both oriented the same
way, \textit{e.g.} towards the hyperbola; they therefore induce
opposite kinks. We call such a departure from the perfect
ellipse, distorted by off-plane kinks, a 'turtle'. One can
eventually imagine elementary kinks of the sort, all of the same
sign, having a high density on the ellipse and continuously
tilting its plane. Such tilted ellipses have been observed in
8CB and 9CB \cite{meyer1}. The situation observed in Fig.8a
results from the presence of a quasi planar \textit{pretilted}
anchoring. A unique direction of pretilt is in conflict with the
presence of an entire ellipse parallel to the boundary in its
close vicinity; hence opposite displacements of different parts
of the ellipse along the vertical direction, such that one part
gets off the boundary, and is virtual. Fig.8c illustrates a
double-kinked ellipse of a turtle type observed from the side in
a thick ($\approx100\mu m$) 8OCB sample.\\
\noindent We remark that both kinking modes, as depicted in Fig.
7, and 8, can be found deep in the smectic A phase.

\subsection{On the origin of deviations from Darboux's theorem}

The just alluded kinking processes can bring large deviations to
Darboux's law; reciprocally it is clear that the deviations from
Darboux's law mean a modification of the shape of the ideal FCD
conics, i.e. the presence of kinks (at the scale of the layers,
since they are not visible with the optical microscope) and of
their attached dislocations.  These dislocations necessarily
disperse through the medium, outside and/or inside the FCD.
Infinitesimal dislocations, if alone, would result, as stated
above, in an extra curvature of the layers. Two cases arise:
either the deformed layers keep parallel, hence the layer normals
keep straight, and one gets eventually a new ideal FCD, or there
is a deviation to straightness of the layer normals, and
consequently a layer thickness variation (this case falls within
the province of the Kroener's dislocation densities
\cite{kroener}), \textit{i.e.} a process of high energy if not
relaxed, at least in part, by finite edge dislocations. It
suffices then to consider only those latter. The edge components
of the attached dislocations that are dispersed \textit{inside}
the FCD break the parallelism of the inside layers. The congruence
of the layer normals is thus no longer a set of straight lines.
This is another way of explaining the variation to Darboux's
theorem. This could have been stated from the start: \textit{edge
dislocation densities break Darboux's theorem, because they break
the layer parallelism}. But this statement comprehends deviations
to Darboux's theorem that are more general than those where the
focal manifolds are degenerate to lines. The focal manifolds of a
congruence of curved normals are generically 2D surfaces, not
lines. We see that the fact that these surfaces are degenerate
into lines comes from the \textit{attachment} of the dislocations
in question to the original focal lines. To conclude, the
occurrence of deviations to Darboux's theorem for a set of focal
\textit{lines} means that the conics are (densely) kinked and
dislocations attached to those kinks.

\subsection{The kinked (split) hyperbola}

The shape of the layers is cylindrical about the central zone of
the hyperbola, near its apex (which is also the physical focus of
the ellipse). But the layers are practically perpendicular to the
hyperbola at a distance from the plane of the ellipse of order
$a$; the wedge disclination smoothly vanishes far from the ellipse
plane.  In between, the layers display cusps, the lesser
pronounced the more distant from the ellipse.  Hyperbolae are
lines of easy coalescence of screw dislocations, as observed long
ago \cite{williams2}.

The presence of kinks on the hyperbola is a delicate matter;
because it is a $k=1$ wedge disclination ($\Omega= 2\pi$), Eq.
(\ref{bdislAB}) and (\ref{dbdislds}) do not apply directly. A way
of solving the question is to consider that the line is made of
two $k=\displaystyle \frac{1}{2}$ lines, indicating that
dislocations with total Burgers vectors twice as large can attach
to a kink of the same size as in the $k=\displaystyle \frac{1}{2}$
case.

Another situation is worth considering. In incomplete focal
domains of the type represented Fig.9b (called \textit{fragmented}
domains), the hyperbola belongs to the boundary of the domain.  It
is then no longer a $k=1$ disclination but a $k=\displaystyle
\frac{1}{2}$ disclination, as if it were split all along its
length. Fragmented FCDs , noted fFCD for short, and already
recognized by G. Friedel \cite{friedelb}, are easily obtained in a
confined sample with degenerate boundary conditions. A fFCD is
bound by a segment of the ellipse and by a segment of the
hyperbola, and four fragments of cones of revolution. Thus both
segments are $k=\displaystyle \frac{1}{2}$ disclination line
segments.

As a consequence, fFCDs are generally aligned, attached by the
ends of the disclination segments, such attachments being required
by the conservation of the disclination strength.  But observe
that a hyperbola H (resp. an ellipse E) can be attached
indifferently either to another H (resp. an E) or to an E (resp. a
H).

One can imagine that the ellipse $E_1$ of a $FCD_1$ is attached to
$H_2$ of a $FCD_2$, while the hyperbola $H_1$ of the $FCD_1$ is
attached to $E_2$ of the $FCD_2$. Such a set of line segments
attached by their extremities is topologically equivalent to a
double helix.  This geometry, with sequences of the ...HEHEH...
type, was observed long ago by C. E. Williams \cite{williams1} at
the N$\longrightarrow$Sm transition; it is at the origin of
helical giant screw dislocations. Fig. 10a shows an elementary
fFCD having the shape of a tetrahedron; the disclination segments
are of opposite 'concavities', which implies that the ellipse
segment is chosen close to the physical focus. Such tetrahedra are
documented in [10].  Fig. 10b shows the abuting of several
tetrahedra, which are no longer perfect fFCD volumes.\\

Let us also mention the observation, also reported in
\cite{meyer1}, of a mobile kink (several microns long)
perpendicular to the $k=\displaystyle \frac{1}{2}$ hyperbola of a
fFCD, moving in the direction of the physical focus, but nucleated
far from it, at a distance large compared to $a$. There is no
doubt that dislocations, dragged along the hyperbola, are attached
to this mobile kink; their Burgers vectors, that are perpendicular
to the layers, are practically parallel to the asymptotic
direction of the hyperbola, at a distance from the ellipse plane,
which indicates that they are of screw character. This might be an
indication of a mechanism by which screw dislocations align along
a (split) hyperbola.

\subsection{Focal Conic Domains at the Sm $\longrightarrow N$ transition}

FCDs that are immersed in the bulk (they are of the type
represented Fig.9a, and generally gather into tilt boundaries)
disappear rather suddenly about $0.5^{\circ}C$ before the
transition, by an \underline{instability} mechanism  that might
imply a sudden multiplication of dislocations. The spontaneous
multiplication of screw dislocations close to the SmA
$\longrightarrow$ N transition is a well documented fact in
\textit{lyotropic} systems \cite{allaina,dhez}, which inclines us
to believe that the phenomenon of spontaneous multiplication of
dislocations (screw but also edge) is very general. The capture of
free edge dislocations by the ellipse modifies its geometric
features $e$ and $a$, Fig.11. Free dislocations of the same
(\textit{resp.} opposite) sign as the dislocations attached to the
ellipse, if captured, would increase (\textit{resp.} decrease) its
size ($2a \longrightarrow 2a+b_{disl}$), either at $e$ constant
(then the asymptotic directions stay constant), or not. Boundary
conditions play a dominant role in this relaxation process. Notice
that, after a possible increase in size, the ellipses eventually
always decrease in size when the temperature increases, the
smallest ellipses disappearing first. For the ellipses belonging
to a tilt grain boundary, this implies that the boundary area
occupied by dislocations (the so-called residual boundary)
increases with temperature. This is in agreement with the model
developed in \cite{kleman1}, which relates the residual boundary
to the material constants; in particular a decrease of the
compression modulus $B$ must result in an increase of the residual
area.

\section{CONCLUSIONS}

This paper investigates from a theoretical point of view some
features of the FCD transformations that have been observed, in
the smectic phase, when approaching the nematic phase. These
very spectacular phenomena happen in a large temperature domain
($\Delta T=T_{AN}-T^* \approx$ half a degree in 8CB, which is
the chemical we used for quantitative observations; the other
compounds yield qualitatively equivalent results) in which it is
believed that the variations of the material constant $B$ are
large enough to allow significant variations of the dislocation
line energy and the multiplication of fresh dislocations.  At
the same time $K_1$ and also $\bar{K}$ (as we assume) do not
vary in comparable proportion, so that the energy of focal conic
domains is not appreciably changed.

We have tried to discuss the general principles at the origin of
these transformations that are due to the direct interaction
between FCDs and finite Burgers vector dislocations. There is no
doubt that infinitesimally small Burgers vector dislocations are
also playing a role, in particular in the phenomena of viscous
relaxation \cite{klemanbook,friedel3}, but this is not discussed.
The general principles that we advance are geometrical and
topological in essence.  The mechanisms that obey these principles
seem to be plenty, depending in particular on the boundary
conditions and the precise FCD texture.  The examples we have
given are few, and are chosen for the sake of illustration.

The SmA $\longrightarrow$ N transition is one of the most
debated liquid crystal phase transitions
\cite{degennes,tonera,tonerb,helfrich}. This is not the place to
enter into the detail of this debate, inasmuch as our results,
even if they stress the importance of defect interplays in the
critical region, are not directly related to the very proximity
of the transition, which has been examined by several authors
with great accuracy (\textit{e.g.} \cite{yethiraj}).

The question that is at stake is rather why the interactions occur
at temperatures definitively lower than $T_{AN}$ and result in an
instability of the FCDs. More details about the instability will
be given in a forthcoming publication.

\section{APPENDIX}

We envision a curved disclination line $\cal L$, carrying a rotation vector
$\overrightarrow{\Omega}$ constant in length and in direction.  Let $P$ be a
point on the cut surface bound by $\cal L$.

We first assume that $\overrightarrow{\Omega}$ is attached to some
well-defined point $O$ (Fig.12). The relative displacement of the
two lips of the cut surface at $P$ is:

\begin{equation}
\overrightarrow{d}_{P}(O)=\overrightarrow{\Omega} \times \overrightarrow{OP}
\end{equation}
%eq. 7

which is large on the line $\cal L$ if $P$ is taken at some point
$M$ on $\cal L$. Consequently in the generic case $\cal L$(0,
$\Omega)$ has a very large core singularity, thus large
accompanying stresses and a large core energy. On the other hand
the cut surface displacement vanishes at $M$ if
$\overrightarrow{\Omega}$ is attached to $\cal L$ at $M$, but then
it does not vanish at $N=M+\overrightarrow{dM}$. There is still a
large core singularity along $\cal L$, except at $M$. The Volterra
process, when applied in its standard form, does not provide a
solution to the construction of a curved disclination with well
relaxed stresses.

An extended conception of the Volterra process solves the problem.
Assume that there is a copy of the rotation vector
$\overrightarrow{\Omega}$ attached to all the points of $\cal L$,
and consider the effect of such rotation vectors on a point $P$
belonging to the cut surface. We have, for each other $M$
belonging to $\cal L$, a value of the relative displacement of the
lips of the cut surface which can be written:

\begin{equation}
\overrightarrow{d}_{P}(M)=\overrightarrow{\Omega} \times \overrightarrow{MP}
\end{equation}
%eq. 8

Each M on $\cal L$ yields another value of the relative
displacement at the same point P of the cut surface, but this
difficulty can be solved by the introduction of a set of
\textit{infinitesimal} dislocations attached to the disclination
line all along $\cal L$, Fig. 13.  Indeed, let M and N = M + dM be
two infinitesimally close points on $\cal L$.  We have:

\begin{equation}
d_{P}(\overrightarrow{
M}+d\overrightarrow{M})-d_\textbf{P}(\overrightarrow{M})=\overrightarrow{\Omega}
\times d\overrightarrow{M}
\end{equation}
%eq. 9

which is independent of $P$. The quantity $d\overrightarrow{b(M)}
=\overrightarrow{\Omega} \times \overrightarrow{dM}$ is the infinitesimal
Burgers vector of the infinitesimal dislocation attached to $\cal L$ at point
$M$ \cite{friedel1}.

The above equations are established for a small angle of rotation
vector $|\overrightarrow{\Omega}|$.  In the general case
$\overrightarrow{\Omega}$ has to be replaced by $\displaystyle
\frac{1}{2} \sin \frac{\Omega}{2}\overrightarrow{t}$, where
$\overrightarrow{\Omega}=\Omega \overrightarrow{t}$.

\acknowledgments We acknowledge fruitful discussions with Dr. V.
Dmitrienko and Prof. J. Friedel. We are grateful to Dr. J.-F.
Blach for providing us with glass plates treated for special
anchoring conditions.

\begin{center}
Figure Captions
\end{center}

Fig.1: (Color on line) a) Complete FCD with negative Gaussian
curvature Dupin cyclides, sitting inside cylinders of revolution
meeting on the ellipse.  The cyclides cross the ellipse plane at
right angles; their intersections with the ellipse and the
hyperbola, when they exist, are conical points. b) Dupin cyclides
fragments with positive and negative Gaussian curvature, so chosen
that the ellipse is still singular but the hyperbola has no
physical realization.  A FCD  with positive and negative Gaussian
curvature both present, the hyperbola singular and the ellipse not
physically realized, is illustrated in \cite{klemanbook}.

Fig.2: (Color on line) In a ideal FCD the ellipse and the
hyperbola project orthogonally along two conics which intersect at
right angles. (Photographs longer edges  $\approx 200\mu m$): a)
8OCB, between two untreated glass substrates, sample thickness
($\approx 100\mu m$), $7 ^{o}C$ below the transition, polarized
light microscopy; Darboux's theorem obeyed; the FCDs with parallel
hyperbola asymptotes form a tilt boundary of the type schematized
in Fig.3; b) 8CB, $0.5 ^{o}C$ below the transition, polarized
light microscopy; Darboux's theorem disobeyed as demonstrated in
the lower photograph: the solid lines are tangents to the
disclinations and the dashed lines perpendicular to them; a very
visible deviation from the Darboux's theorem is encircled on the
upper photograph.

Fig.3: (Color on line) Tilt boundary split into FCDs. TOP:
schematic, adapted from \cite{klemanbook}; BOTTOM: 8CB, polarized
light microscopy; the tilt boundary is seen edge-on; the edge of
the photograph $\approx 100\mu m$ long.

Fig.4: F, the physical focus, is the center of the (circular)
intersections of the layers with the plane of the ellipse,
inside the ellipse; $\overrightarrow{t}$ is a unit vector along
the local rotation vector; the $k=\displaystyle \frac{1}{2}$
disclination ellipse is of mixed (twist-wedge) character all
along, except at the ends of the major axis, where it is wedge.

Fig.5: Kink on a wedge disclination line, see text.

Fig.6: The ellipse in polar coordinates.  The radius of
curvature of the circle centered in the focus F and tangent to
the apex is $a-c$, which is smaller than the radius of curvature
$\displaystyle \frac{b^2}{a}$ of the ellipse at the apex. This
circle is thus entirely inside the ellipse.  All the circles and
the arcs of circles of the figure are centered in F. They figure
intersections of the smectic layers with the plane of the
ellipse.

Fig.7: Double kinks with a dislocation \textit{outside} the FCD;
a)- Mouse patterns in 8CB, free standing film, rim region; the
thickness decreases downward; longer side of the photograph
 $\approx 200\mu m$); b)- model.

Fig.8: Views of a double kink with a dislocation \textit{inside}
the FCD (longer side of photographs $\approx 200\mu m$): a)-
turtle patterns in 8CB, demonstrating that the ellipses are
divided into two parts  not located at the same level, the screw
dislocations attached to the kinks are visible; b)- model of a
double kink linked by a unique dislocation located inside the FCD;
c)- a double kinked ellipse (kinks are shown by arrows) of the
turtle type observed from the side (8OCB in a gap of the thickness
$\approx 100\mu m$ between two untreated glass substrates).

Fig.9: (Color on line) Incomplete FCDs. a)- FCD bound by two cones
of revolution meeting on the ellipse, with apices at the
terminations of the hyperbola segment; b)- A hyperbola-split
fragmented FCD (fFCD). The fFCD is bound by i) two fragments of
cones of revolution with apices at the terminations of the
hyperbola segment and limited to the ellipse segment, ii) two
fragments of cones of revolution with apices at the terminations
of the ellipse segment and limited to the hyperbola segment. The
director field on the boundaries is indicated, not the cyclide
intersections with these boundaries.

Fig.10: (Color on line) a)- elementary tetrahedra fFCD liable to
form a portion of a double helix due to the favourable
concavities; b)- abuting of several tetrahedra.

Fig.11: Edge dislocations mobile in the plane of a perfect ellipse
(belonging to a ideal FCD) and attaching to it. The consecutive
modification of the FCD results from a relaxation process towards
a new ideal FCD; this process has to respect the boundary
conditions, \textit{e.g.} $e=const$, if the angle $\omega$ of
misorientation is fixed.

Fig.12: The classic Volterra process for a rotation vector
$\overrightarrow{\Omega}$ attached to $O$. At a point $P$ on the
cut surface, the lips of the cut surface suffer a relative
displacement $\overrightarrow{d}_P(O) = \overrightarrow{\Omega}
\times \overrightarrow{OP}$.

Fig.13: The extended Volterra process for a rotation vector
$\overrightarrow{\Omega}$ attached locally to each point on
$\cal L$. Infinitesimal dislocations are attached all along the
disclination line.


\begin{references}

\bibitem{meyer1}
C. Meyer and M. Kleman, ILCC20, Mol. Cryst. and Liq. Cryst.,
\textbf{437}, 111[1355] (2005).
%1

\bibitem{yuriy}
Yu. A. Nastishin, C. Meyer and M. Kleman, in preparation.
%2

\bibitem{degennes}
P. G. De Gennes and J. Prost, \textit{The Physics of Liquid
Crystals}, second edition, Clarendon Press (1993).
%3

\bibitem{benzekri} M. Benzekri, T. Claverie, J.-P. Marcerou,
and J.-C. Rouillon, Phys. Rev. Lett. \textbf{68}, 16 (1992).
%4

\bibitem{rogez} D. Rogez, D. Collin and P. Martinoty, Eur. Phys. J. E \textbf{14}, 43-47
(2004)
%5

\bibitem{beaubois} F. Beaubois, T. Claverie, J.-P. Marcerou, J.-C. Rouillon,
and H.-T. Nguyen, C.W. Garland and H. Haga, Phys. Rev. E \textbf{56}, 5 (1997).
%6

\bibitem{yethiraj} A. Yethiraj, J. Bechhoefer, Phys. Rev. Lett. \textbf{84}, 16 (2000).
%7

\bibitem{barbero}
G. Barbero and V. M. Pergamenshchik, Phys. Rev. E \textbf{66} 051706 (2002).
%8

\bibitem{friedela} G. Friedel et F. Grandjean, Bull. Soc. Fr. Min\'{e}r. \textbf{33}, 192, 409 (1910).
%9

\bibitem{friedelb}G. Friedel, Ann. Phys. (Paris) \textbf{18}, 273 (1922).
%10

\bibitem{klemanbook}
M. Kleman and O. D. Lavrentovich, \textit {Soft Matter Physics.
An Introduction}, Springer N.Y. (2003).
%11

\bibitem{williams0}
C. E. Williams, \textit{D\'{e}fauts de structure dans les
smectiques A}, PhD Thesis, Orsay (France) (1976).
%12

\bibitem{fournier}
J.-B. Fournier and G. Durand, J. Phys. II France \textbf{1}, 845
(1991).
%13

\bibitem{maurice}
M. Kleman and O. D. Lavrentovich, Phys. Rev. E\textbf{61}, 1574 (2000).
%14

\bibitem{boltenhagen1}
P. Boltenhagen, O. Lavrentovich and M. Kl\'{e}man, Phys. Rev. A
\textbf{46}, 1743(1992).
%15

\bibitem{williams1} C.E. Williams, Philos. Mag. \textbf{32}, 313 (1975).
%16

\bibitem{kln}
M. Kleman, O. D. Lavrentovich and Yu. A. Nastishin, in Dislocations in Solids,
F. R. N. Nabarro and J. P. Hirth, eds, North-Holland, Amsterdam vol.
\textbf{12}, 147 (2004).
%17

\bibitem{darboux} G. Darboux, \textit{Le\c{c}ons sur la th\'{e}orie g\'{e}n\'{e}rale des surfaces}, seconde partie,
Gauthier-Villars, Paris (1914).
%18

\bibitem{kleman1} M. Kleman and O. D. Lavrentovich, Eur. Phys. J. E \textbf{2}, 47 (2000).
%19

\bibitem{bourdon} L. Bourdon, M. Kl\'{e}man and J. Sommeria, J. de Physique \textbf{43}, 77 (1982).
%20

\bibitem{boltenhagen2} P. Boltenhagen, O. D. Lavrentovich and M. Kl\'{e}man, J. Phys. II France
\textbf{1}, 1233 (1991).
%21

\bibitem{frank} F. C. Frank, Discuss. Faraday Soc. \textbf{25}, 19 (1958).
%22

\bibitem{friedel1} J. Friedel and M. Kl\'{e}man, J. de Phys. \textbf{30}, C4:43 (1969).
%23

\bibitem{kroener} E. Kroener, in Physics of Defects, Les Houches 1980 Session XXXV (eds. R.
Balian, M. Kl\'{e}man and J.P. Poirier), North-Holland, Amsterdam, (1981).

\bibitem{williams2} C. E. Williams and M. Kl\'{e}man, Philos. Mag. \textbf{33}, 213 (1976).

\bibitem{allaina} M. Allain and J.-M. diMeglio, Mol. Cryst. Liq. Cryst. \textbf{124}, 115
(1985), Europhys. Lett. \textbf{2}, 597 (1986).

\bibitem{dhez}
O. Dhez, S. K\"{o}nig, D. Roux, F. Nallet and O. Diat, Eur. Phys.
J. E\textbf{3}, 377 (2000).

\bibitem{friedel3} J. Friedel, private communication.

\bibitem{tonera} J. Toner and D. Nelson, Phys. Rev. B\textbf{24}, 363 (1981).

\bibitem{tonerb} J. Toner, Phys. Rev. B\textbf{26}, 1 (1982).

\bibitem{helfrich} W. Helfrich, J. de Physique \textbf{39}, 1199 (1978).



\end{references}
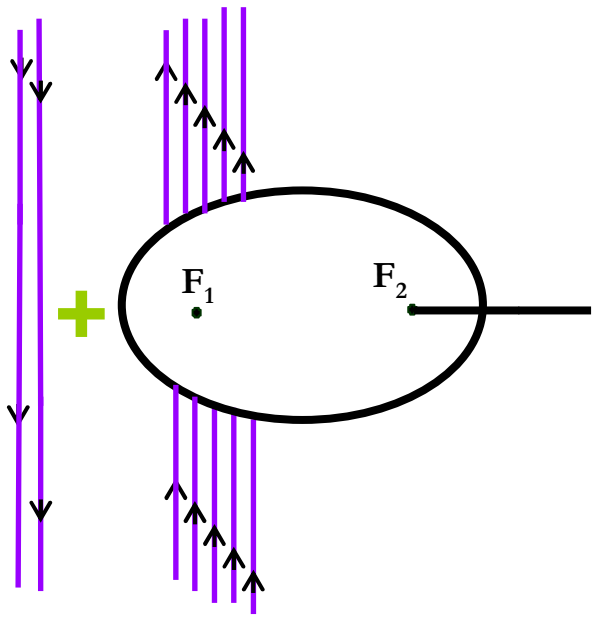
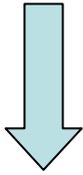
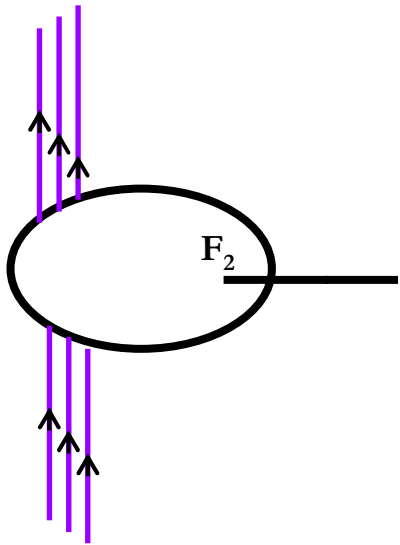
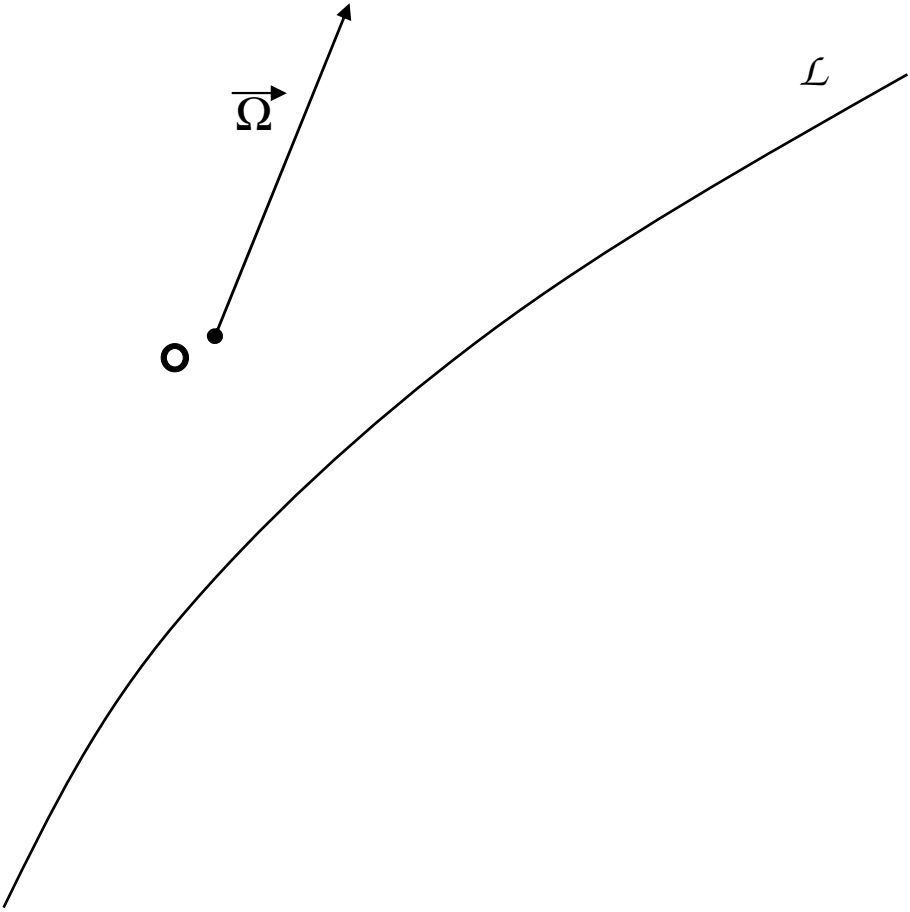
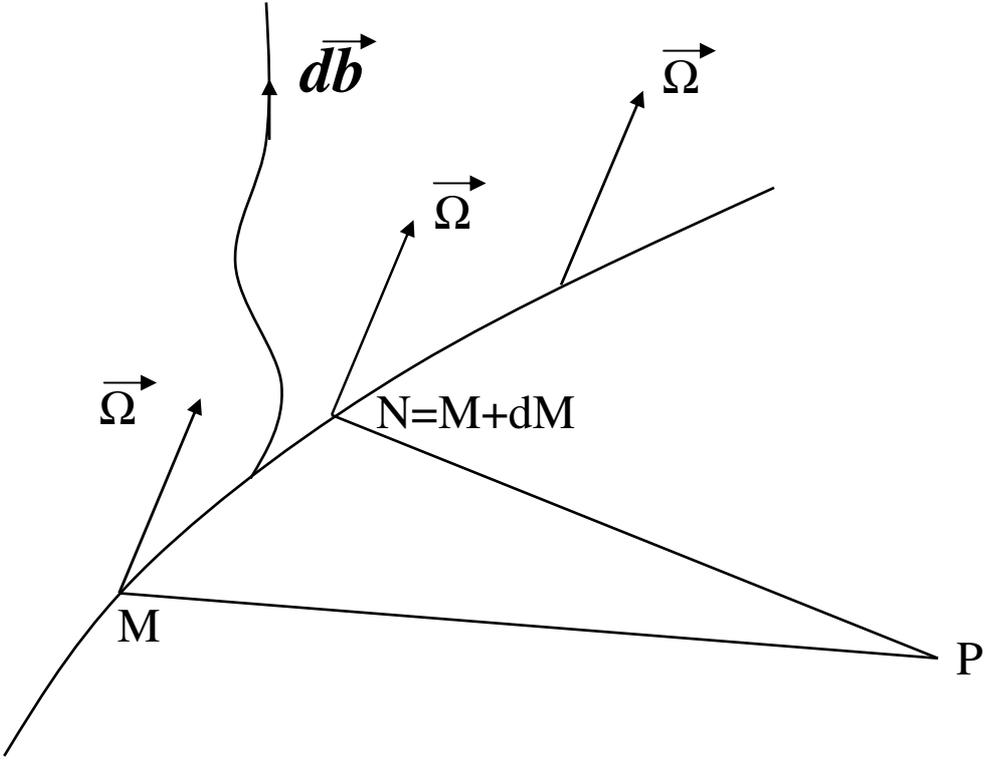
\end{document}